\documentclass[twocolumn]{revtex4}
\usepackage{graphicx}
\begin{document}

\title{Modified gravity with $f(R) = \sqrt{R^2 - R_0^2}$}
\author{Shant Baghram, Marzieh Farhang, Sohrab Rahvar
\vspace{0.15cm}} \affiliation{\it Department of Physics, Sharif
University of Technology, P.O.Box 11365-9161, Tehran, Iran}

\begin{abstract}
Here in this work we propose a modified gravity with the action of
$f(R) = \sqrt{R^2 - R_0^2}$ instead of Einstein-Hilbert action to
describe the late time acceleration of the universe. We obtain the
equation of the modified gravity both in the metric and Palatini
formalisms. An asymptotic solution of gravity equations
corresponding to a constant Ricci scalar causes a late time
acceleration of the universe. We do a conformal transformation in
the action of the modified gravity and obtain the equivalent
minimally coupled scalar-tensor gravity. The equivalent Brans-Dicke
gravity of this model is also studied. To examine this model with
the observation, the perihelion Precession of the Mercury is
compared with our prediction and we put an upper constraint of
$R_0<H_0^2$. This range for $R_0$ is also in agreement with the
cosmological acceleration at the present time. Finally we show that
this action has instability for the small perturbations of the
metric in vacuum solution in which adding a quadratic term of Ricci
scalar can stabilize it.
\end{abstract}


\maketitle

\section{Introduction}
Recent cosmological data from the Supernova Type Ia (SNIa) and
Cosmic Microwave Background (CMB)\cite{Ben03,Net01,Hal02} indicate
that universe is currently in the accelerating phase of expansion.
The effect of acceleration of the universe on the Supernova Type Ia
(SNIa) is making their apparent luminosity dimmer than our
expectation \cite{R04,Perl99,Tor03}. The cosmological constant can
explain this observational effect. The best fit of the data with the
standard (Friedmann Robertson Walker) FRW model results an energy
density of $10^{-47} Gev^{4}$ for the cosmological constant. This
value is a 55 order of magnitude smaller than the energy density of
vacuum in quantum field theory (for review
see\cite{Car100,Peeb103}). The cosmological constant is a
geometrical term, however it can be regarded as a perfect fluid with
the equation of state of $w=-1$. This equation of state provides a
constant density during the expansion of the universe. To have the
present universe the cosmological constant should be fine-tuned and
a small change in this term induces a dramatic change in the destiny
of universe.

To solve the cosmological constant problem, the Quintessence model
as a time varying  cosmological constant due to slow rolling of a
scalar field is proposed. This model can cure the fine-tuning
problem at the early universe, however there is no physical
candidate for the origin of this scalar field
\cite{peb88,wet88,sah99,arm00,pad03,cal02,cal03,w04,mov06,Arm99,Mers01,carTr03,par99}.
The other alternative approach dealing with the acceleration problem
of the universe is changing the gravity law through the modification
of action of gravity by means of using $f(R)$ instead of the
Einstein-Hilbert action. Some of these models as $1/R$ and
logarithmic models provide an acceleration for the universe at the
present time
\cite{bar05,Noj03,Noj103,Def02,Fre02,Ahm02,Ark02,Dva03}.

The back-reaction of the structures at large scales on the dynamics
of the background of the universe \cite{Bran03} is also proposed as
a candidate for interpretation of the acceleration of the universe.
Structured FRW universe is another approach which provides a larger
luminosity distance in the expanding universe, making the luminosity
of SNIa dimmer \cite{Mans05}. Here in this work we propose a model
for the modified gravity with the action of $f(R) = (R^2 -
R_0^2)^{1/2}$ to provide a positive acceleration for the universe at
the later times of the cosmic history. In section \ref{dynamics} we
derive the modified Einstein equations both in the metric and
Palatini formalisms and show that the solution of maximally
symmetric static universe could be a de-Sitter space. In section
\ref{equi} we show the equivalence of the modified gravity with the
scalar tensor models such as the the Brans-Dicke and non minimally
coupled conformal transformed scalar tensor theories. In Section
\ref{instability} we show the instability of this action for small
perturbations of curvature around the vacuum solution and show that
adding an $R^2$ term can stabilize this model. The conclusions are
presented in Section \ref{conclusion}
\section{Field equation of $f(R)=\sqrt{R^2-R_0^2}$ in Modified Gravity }
\label{dynamics} For an arbitrary action of the gravity there are
two main approaches to extract the field equations. The first one is
the so-called "metric formalism" in which the variation of action is
performed with respect to the metric. In the second approach,
"Palatini formalism", the connection and metric are considered
independent of each other and we have to do variation for those two
parameters independently. Here we obtain the field equations in the
two mentioned approaches:
\subsection{metric formalism}
A generalized form of the action due to geometrical part  of gravity
can be written as follows:
\begin{equation}
S = \frac{1}{16\pi G}\int f(R)\sqrt{-g}dx^4 +S_{matter},
\label{generalized}
\end{equation}
where in the case of $f(R)=R$ we have the ordinary Einstein-Hilbert
gravity. Variation of the action with respect to the metric results
in the field equations as:
\begin{equation}
f'(R)R_{\mu \nu}-\frac{1}{2}f(R)g_{\mu
\nu}-(\nabla_{\mu}\nabla_{\nu}-g_{\mu
\nu}\nabla_{\alpha}\nabla^{\alpha}) f'(R)=8\pi G T_{\mu \nu}.
\label{m_vari}
\end{equation}
Here we choose the geometric part of the Lagrangian as
$f(R)=\sqrt{R^2-R_0^2}$ which implies the field equations as:
\begin{eqnarray}
\label{feq}& &\frac{R
R_{\mu\nu}}{\sqrt{R^2-R_{0}^2}}-\frac{1}{2}\sqrt{R^2-R_0^2}g_{\mu\nu}\\\nonumber
&&-\frac{R_0^2}{(R^2-R_0^2)^{3/2}}[\nabla_{\mu}\nabla_{\nu}R-
g_{\mu\nu}{\nabla_{\alpha}\nabla^{\alpha}R}\\\nonumber&&
-\frac{3R}{R^2-R_0^2}(\nabla_{\mu}R\nabla_{\nu}R-g_{\mu\nu}{\nabla^{\alpha}R\nabla_{\alpha}R})]=8\pi
G T_{\mu \nu}.
\end{eqnarray}
For the special case of $R_0=0$ we will recover the familiar
Einstein field equations. One of the features of this action is that
we will have an intrinsic minimum curvature for the space. Taking
the trace of (\ref{feq}) results:
\begin{eqnarray}
&&\frac{R^2}{\sqrt{R^2-R_0^2}}-2\sqrt{R^2-R_0^2}+\frac{R_0^2}{(R^2-R_0^2)^\frac{3}{2}}[3\nabla_{\alpha}\nabla^{\alpha}R\nonumber\\
&&-\frac{9R}{R^2-R_0^2}\nabla_{\mu}R\nabla^{\mu}R]=kT.
\label{trace1}
\end{eqnarray}
As a special solution we consider a maximally symmetric space-time
without the energy momentum source where $R$ is independent of space
and time. This condition requires $R=\sqrt{2}R_0$ and it can happen
at the ultimate stage of the expansion of the universe. Now we solve
the field equations for the modified gravity in the two special
cases of static spherically symmetric and spatially homogenous
spaces:
\subsubsection{Spherical static space}
We take a spherically symmetric Schwarzschild-like metric as:
\begin{equation}
ds^2 = -e^{-\lambda(r)}dt^2 + e^{\lambda(r)}dr^2 + r^2(d\theta^2 +
\sin(\theta)^2d\phi^2)
\end{equation}
It is just a straightforward task to find the Christoffel symbols
related to this metric. The non-vanishing components of Ricci tensor
are as following:
\begin{equation}
R_{11}=\frac{\lambda^{''}(r)}{2}-\frac{{\lambda^{'2}(r)}}{2}+\frac{\lambda^{'}(r)}{r},
\end {equation}

\begin{equation}
R_{22}=-e^{-\lambda(r)}[1-r\lambda^{'}(r)]+1=\frac{R_{33}}{\sin{\theta}},
\end{equation}
\begin{equation}
R_{44}=e^{-\lambda(r)}(-\frac{\lambda^{''}(r)}{2}+\frac{\lambda^{'2}(r)}{2}-\frac{\lambda^{'}(r)}{r}),
\end{equation}
where the Ricci scalar obtain as:
\begin{equation}
R=e^{-\lambda(r)}(\lambda''-{\lambda'}^2+\frac{4\lambda'}{r}-\frac{2}{r^2})+\frac{2}{r^2}
\label{Ricci}
\end{equation}
By substituting Ricci scalar and tensor in equation (\ref{feq}) we
will have a forth order non-linear differential equation for
$\lambda(r)$ with no analytical solution. A simple case is the
vacuum solution which results in:
\begin{equation}
R=\sqrt{2}R_0
\end{equation}
Substituting in equation (\ref{Ricci}) we obtain $\lambda(r)$ as:
\begin{equation}
\lambda(r)=-\ln(1+\frac{c_1}{r}-\frac{c_2}{r^2}-\frac{\sqrt{2}}{12}R_0r^2)
\end{equation}
In analogy with the Schwarzschild metric in the Newtonian limit we
set $c_1=-\frac{2GM}{c^2}$ and $c_2=0$ and we have the metric as
follows:
\begin{eqnarray}
&&ds^2=-(1-\frac{2GM}{c^2r}-\frac{\sqrt{2}}{12}R_0r^2)c^2dt^2+\nonumber\\&&(1-\frac{2GM}{c^2r}+
\frac{\sqrt{2}}{12}R_0r^2)^{-1}dr^2+r^2(d\theta^2+\sin{\theta}^2d\varphi^2).\\\nonumber
\end{eqnarray}
This metric in the generalized gravity is similar to the
Schwarzschild--de Sitter space in the Einstein-Hilbert action where
$R_{0}=2\sqrt{2}\Lambda$ plays the role of the cosmological
constant. A the solution of generalized Einstein equation for a
perturbation around the vacuum solution in the spherically symmetric
metric is given in Appendix.

There is a list of the observations from the solar system to the
cosmological scales the modified gravity can be examined. Here we
use the perihelion of the Mercury to check the behavior of the orbit
in this model and put constraint on $R_0$ as the parameter of the
generalized gravity. Mercury is the inner most of the four
terrestrial planets in Solar system, moving with high velocity in
Sun's gravitational field. That is why mercury offers unique
possibilities for testing general relativity and exploring the
limits of alternative theories of gravitation. The observed advance
of the perihelion of Mercury that is unexplained by Newtonian
planetary perturbations or solar oblately is \cite{Pir}:
\begin{eqnarray}
                 &&\Delta\omega_{obs}=42.980\pm0.002 \hspace{0.5cm}arcsecond/century \\
                 \nonumber
                 && =2\pi(7.98734\pm0.00037)\times10^{-8}
                 radians/revolution
\end{eqnarray}
The calculation of the perihelion of Mercury in the
Schwarzschild--de Sitter metric has been studied so far in
\cite{Krani}. For the time like geodesics they used Jacobi's
inversion problem and found the constraint of $\Lambda<10^{-55}
cm^{-2}$. From the spherical solution we have the similar constraint
of $R_0<10^{-55} cm^{-2}$ in our model. Comparing with the horizon
size of universe at the present time implies $R_0^{1/2}<H_0$.
\subsubsection{cosmological solution}
\begin{figure}
\includegraphics[height=3in, width= 3.5in, angle=0]{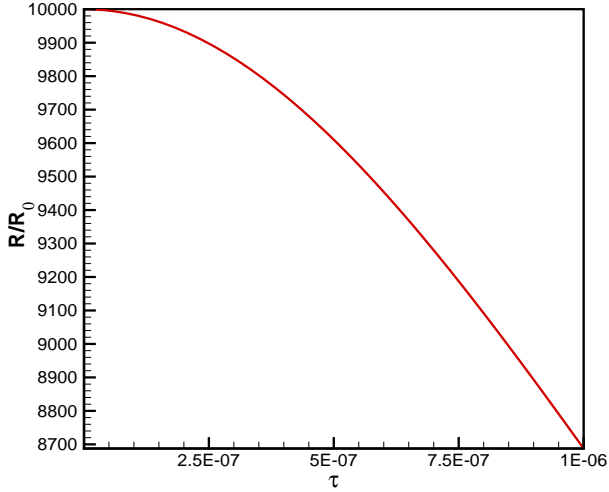}
\caption{Dynamics of $y=R/R_0$ in terms of $\tau = t R_0^{1/2}$ for
the radiation dominate epoch. This dynamics results from the metric
formalism for the field equation. The action for the modified
gravity is taken as $f(R) = \sqrt{R^2 - R_0^2}$.} \label{fig1}
\end{figure}

We start the cosmological solution with the dynamics of universe in
the radiation dominant epoch. For the early universe the equation of
state of cosmic fluid is $p=1/3\rho$ and the right hand side of
equation (\ref{trace1}) is zero. The Ricci scalar for the FRW metric
is as follows:
\begin{equation}
3\ddot{R}R_0^2 - \frac{9R_0^2R\dot{R}^2}{R^2 - R_0^2} + R^4
-3R^2R_0^2 + 2R_0^4 = 0.
\label{trace3}
\end{equation}
We divide this equation by $R_0^4$ and rename $y=R/R_0$ and define a
dimensionless variable as $\tau = tR_0^{1/2}$. Equation
(\ref{trace3}) with the new variables can be written as:
\begin{equation}
3y'' - \frac{9y'^2y}{y^2-1} + y^4 -3y^2 + 2=0
\end{equation}
We expect to have Einstein-Hilbert action for the early times of the
universe which means $y>>1$. The numerical solution of this equation
is shown in Fig. \ref{fig1}.
Here the Ricci scalar should be larger than $R_0$ but comparable
with this term. This assumption implies that Ricci scalar should be
smaller than $H^2$ at that time. From the FRW metric Ricci scalar is
related to the Hubble parameter through: $R = 6\dot{H} + 12H^2$.
Considering $R(t)$ as a small and constant perturbation due to the
modified gravity, the differential equation for the scale factor is
$\ddot{a} + \dot{a}^2 = \epsilon a^2$. By a change of variable as
$x=a^2$ and considering the initial condition of $a=0$ at $t=0$, the
solution of the equation yields: $x(t) = A\sinh(\sqrt{\epsilon}t)$.
Expanding this expression around $t=0$ results in $a(t) = C t^{1/2}
+ \sigma t^{5/2}$, where $\sigma$ is a small value.

This deviation of the scale factor from $a\propto t^{1/2}$ depends
on the Ricci scalar at the radiation epoch. Since $R$ and $R_0$ are
in the order of $H_0^{-2}$, we expect that this model does not alter
the dynamics of the early universe from that of Einstein-Hilbert
action.
\subsection{Palatini formalism}
In the Palatini formalism, the connections are considered as
independent variables, different than the Christoffel symbols of the
metric\cite{sot06}. Varying the action with respect to both the
metric and the connections the corresponding field equations are
obtained as:
\begin{equation}
f'(R)R_{\mu\nu}-\frac{1}{2}f(R)g_{\mu\nu}=\kappa
T_{\mu\nu},\label{metric}
\end{equation}
\begin{equation}
\nabla_\lambda(\sqrt{-g}f'(R)g^{\mu\nu})=0, \label{connection}
\end{equation}
in which we have considered the matter action to be independent of
the connections. From equation(\ref{connection}) we can see that the
connections are the Christoffel symbols of the new metric
$h_{\mu\nu}$ where it is conformally related to the original one via
the equation
\begin{equation}
h_{\mu\nu}=f'(R)g_{\mu\nu}.
\end{equation}
Equation (\ref{metric}) shows that in contrast to the metric
variation approach (see equation \ref{m_vari}), the field equations
are second order in this formalism. The trace of field equations in
the Palatini formalism for $f(R)=\sqrt{R^2-R_0^2}$ is:
\begin{equation}
\frac{R^2}{\sqrt{R^2-R_0^2}}-2\sqrt{R^2-R_0^2}=\kappa T.
\label{trace}
\end{equation}
The vacuum solution of this equation $R=\sqrt{2}R_0$, is the same as
in the metric formalism. Through the conformal transformation of the
metric $g_{\mu\nu}$ to $h_{\mu\nu}$, we get the corresponding Ricci
tensor:
\begin{equation}
 \label{ricci1}
R_{00}=-3\frac{\ddot{a}}{a}+\frac{3}{2}f'^{-2}(\partial_{0}
f')^2-\frac{3}{2}f'^{-1}\bar{\nabla}_0\bar{\nabla}_0 f',
\end{equation}
\begin{equation}
\label{ricci2} R_{ij}=[a \ddot{a}+2 \dot{a}^2 +
f'^{-1}\left\{^{\lambda}_{\mu\nu}\right\} \partial_0 f'
+\frac{a^2}{2}f'^{-1}
\bar{\nabla}_0\bar{\nabla}_0 f']\delta_{ij},
\end{equation}
where $\left\{^{\lambda}_{\mu\nu}\right\}$ and $\bar{\nabla}$ are
associated with $g_{\mu\nu}$. Using eqs. (\ref{metric}),
(\ref{ricci1}) and (\ref{ricci2}) we can derive the modified
Friedmann equation:
\begin{equation}
6H^2+6Hf'^{-1}\partial_{0}f'+\frac{3}{2}f'^{-2}(\partial_{0}f')^2=\frac{\kappa(\rho+3p)+f}{f'}.
\label{frw_pal}
\end{equation}
Now we want to obtain the dynamics of universe in the radiation
dominant epoch. For this epoch implying the equation of state $p =
1/3 \rho$, from the equation (\ref{trace}) $T=0$, we get a constant
value of $R_{r} = \sqrt{2}R_0$ for the Ricci scalar. Substituting it
in (\ref{frw_pal}) results in:
\begin{equation}
6H^2=\frac{2\kappa\rho_r+R_0}{\sqrt{2}} \label{radiation}
\end{equation}
in which the radiation density changes by the scale factor of the
universe as $\rho_r=\rho_0a^{-4}$. The solution of equation
(\ref{radiation}) results:
\begin{equation}
a(t)\propto \sinh^{1/2}
(\sqrt[4]{\frac{2R_0^2}{9}}t).
\end{equation}
This solution is similar to the case of metric formalism, showing
that for the action of $f(R) = \sqrt{R^2 - R_0^2}$, we have almost
the same dynamics for the early universe in both Palatini and metric
formalism.

\begin{figure}
\includegraphics[height=3in, width= 3.5in, angle=0]{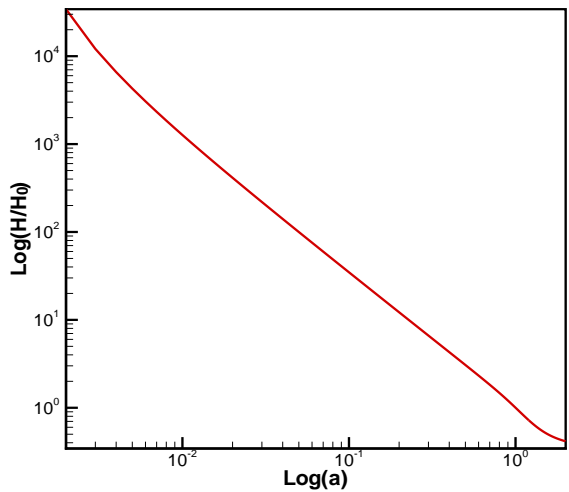}
\caption{Numerical solution of Equations (\ref{pp1}) and
(\ref{pp2}). Here we omit the Ricci scalar between the two equations
and obtain a direct relation between the Hubble parameter and scale
factor in the logarithmic scale.} \label{pal_h_a}
\end{figure}

\begin{figure}
\includegraphics[height=3in, width= 3.5in, angle=0]{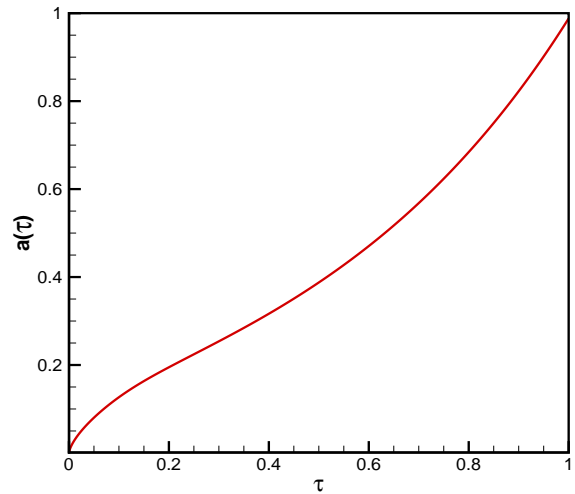}
\caption{Numerical solution of scale factor in terms of cosmic
 time, normalized to the $R_0^{-1/2}$.} \label{a_pal}
\end{figure}

For the matter dominant epoch we use the direct dependence of the
Hubble parameter and scale factor on $f(R)$ as\cite{ama06}:
\begin{eqnarray}\label{pp1}
H^2 & = & \frac{1}{6 f'} \frac{Rf'-3f}
{\left(1-\frac 32 \frac{f''(Rf'-2f)}{f'(Rf''-f')}\right)^2},\\
a & = &
\left(\frac{1}{\kappa\rho_0}\left(Rf'-2f\right)\right)^{-\frac{1}{3}}.
\label{pp2}
\end{eqnarray}
The numerical solution of these equations for $f(R) = \sqrt{R^2
-R_0^2}$ gravity is shown in Fig. \ref{pal_h_a}. Here we obtain the
Hubble parameter normalized to its present value in terms of the
scale factor, in logarithmic scale.
By the numerical solution of H=H(a) a direct relation between the
scale factor and cosmic time normalized to the $R_0^{-1/2}$ is shown
in Fig. \ref{a_pal}.
\section{Equivalence of the Modified gravity with scalar tensor theories}
\label{equi} Here in this section we discuss the equivalence of the
modified gravity with the scalar tensor theories, such as
scalar-tensor coupled to the curvature like Brans-Dicke and
non-minimally coupled one which is a result of conformal
transformation from the Jordan to the Einstein frame\cite{brans}.
Here in this section we study the corresponding theories of
$f(R)=\sqrt{R^2-R_{0}^2}$ action.
\subsection{Brans-Dicke theory}
\begin{figure}
\includegraphics[height=3in, width= 3.5in, angle=0]{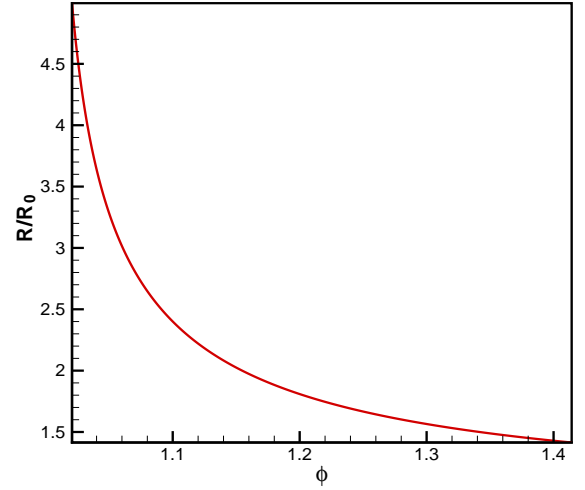}
\caption{The dependence of the scalar field on the Ricci scalar in
Brans-Dicke correspondence to the modified gravity.} \label{phiR}
\end{figure}

Theories in which scalar fields are coupled directly to the
curvature, are termed scalar-tensor gravity \cite{tey83}. Such
theories can be found in the low-energy effective string theory
which couples a dilation field to the Ricci curvature tensor. The
simplest and the well known one is the Brans-Dicke (BD) theory. The
BD theory is a generalization of the general relativity with the
action of:
\begin{eqnarray}
&&S_{BD}=\frac{1}{2\kappa}\int{d^{4}x\sqrt{-g}[\phi{R}-\frac{\omega_{0}}{\phi}(\partial_{\mu}\phi\partial^{\mu}\phi)
-V(\phi)]} \nonumber\\ & &+S_{M}(g_{\mu\nu},\psi), \label{BD}
\end{eqnarray}
where the free parameter $\omega_{0}$ is often called Brans-Dicke
parameter and for the case of $\omega_{0}\rightarrow\infty$ GR is
recovered. The field equation that one derives from the action
(\ref{BD}) by varying with respect to the metric and the scalar
field are
\begin{eqnarray}
\label{f_BD1} &&
G_{\mu\nu}=\frac{\kappa}{\phi}T_{\mu\nu}+\frac{\omega_{0}}{\phi^2}(\nabla_{\mu}\phi\nabla_{\nu}\phi-\frac{1}{2}g_{\mu\nu}\nabla^{\lambda}\phi\nabla_{\lambda}\phi)
\\ \nonumber
&&\\\nonumber
&&+\frac{1}{\phi}(\nabla_{\mu}\nabla_{\nu}\phi-g_{\mu\nu}\nabla_{\alpha}\nabla^{\alpha}\phi)-\frac{V}{2\phi}g_{\mu\nu}
\end{eqnarray}
\begin{equation}
\label{f_BD2}
\frac{2\omega_{0}}{\phi}\nabla_{\alpha}\nabla^{\alpha}{\phi}+R-\frac{\omega_{0}}{\phi^{2}}\nabla_{\alpha}\phi\nabla^{\alpha}\phi-V^\prime=0
\end{equation}
where $G_{\mu\nu}=R_{\mu\nu}-\frac{1}{2}Rg_{\mu\nu}$ is the Einstein
tensor and $T_{\mu\nu}=\frac{-2}{\sqrt{-g}}\frac{\delta
S_{M}}{\delta g^{\mu\nu}}$ is the stress-energy tensor. We take the
trace of Eq.(\ref{f_BD1}) and combine it with Eq. (\ref{f_BD2}) to
omit R, which results in the differential equation for the dynamics
of scalar field as:
\begin{equation}
(2\omega_{0}+3)\nabla_{\alpha}\nabla^{\alpha}\phi=\kappa T+\phi
V^{\prime}-2V
\end{equation}
We can also find an equivalent action for the BD theory with
introducing a modified gravity as $f(R)$. Let us use an auxiliary
field $\chi$ and write an equivalent action as:
\begin{equation}
S=\frac{1}{2\kappa}\int{d^{4}x\sqrt{-g}(f(\chi)+f^{\prime}(\chi)(R-\chi))}+S_{M}(g_{\mu\nu},\psi).
\end{equation}
Varying with respect to $\chi$ leads to the equation $\chi=R$ if
$f^{\prime\prime}(\chi)\neq0$. By redefining the field $\chi$,
$\Phi=f^{\prime}(\chi)$ and setting:
\begin{equation}
V(\Phi)=\chi(\Phi)\Phi-f(\chi(\Phi)) \label{potentialBD}
\end{equation}
the action takes the form of
\begin{equation}
S=\frac{1}{2\kappa}\int{d^{4}x\sqrt{-g}(\Phi{R}-V(\Phi))}+S_{M}(g_{\mu\nu},\psi)
\end{equation}
Comparison with the action in (\ref{BD}) reveals that we have an
equivalent Brans-Dicke theory with $\omega_{0}=0$. In other words
$f(R)$ theories are fully equivalent to a class of Brans-Dicke
theories with vanishing kinetic term. For the case of our action,
$f(R)=\sqrt{R^2-R_{0}^2}$ we can easily find the proper Brans-Dicke
potential using equation (\ref{potentialBD}):
\begin{equation}
V(\Phi)=R_{0}\sqrt{\Phi^2-1}
\end{equation}
The Ricci scalar also depends on the scalar field by (see
Fig.\ref{phiR}):
\begin{equation}
R=V^{\prime}=R_{0}\frac{\Phi}{\sqrt{\Phi^2-1}}.
\end{equation}
For the early universe with large Ricci scalar we have
$\phi\rightarrow 1$ and for the accelerating epoch of the universe,
$\phi\rightarrow \infty$ which results in $R\rightarrow R_0$.
Substituting the potential in the Brans-Dicke action, the final form
of the equivalent action with the modified gravity is:
\begin{equation}
S_{BD}=\frac{1}{2\kappa}\int{d^{4}x\sqrt{-g}(\Phi{R}-R_{0}\sqrt{\Phi^2-1})}+S_{M}(g_{\mu\nu},\psi)
\end{equation}
\subsection{Non-minimally coupled scalar-tensor gravity: conformal transformation}
Metric theories of gravitation which depend on the scalar curvature
in a nonlinear way, are usually called improperly "nonlinear
gravity" (NLG) models \cite{mag93}. A suitable conformal
transformation on metric can change the NLG lagrangian from the
original frame so-called Jordan frame into Einstein-Hilbert one,
minimally coupled to a scalar field. It is therefore claimed that
any NLG theory is mathematically equivalent to General Relativity
(with the scalar field), but from the physical point of view there
is a debate in literature if there is an experiment to distinguish
them \cite{mag93}.

\begin{figure}
\includegraphics[height=3in, width= 3.5in, angle=0]{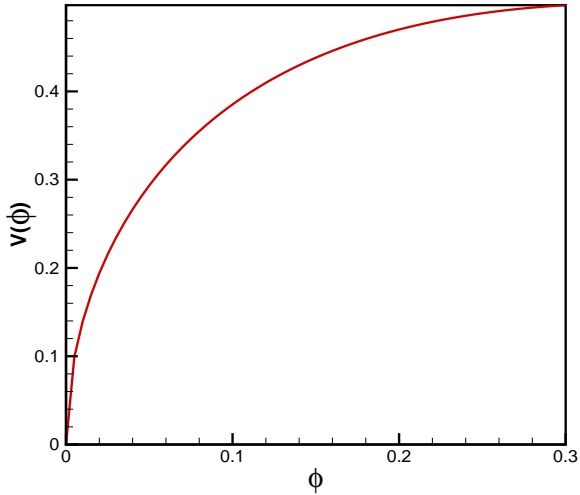}
\caption{Dependence of potential on the scalar field results from
the conformal transformation from Jordan to the Einstein frame in
$f(R) = \sqrt{R^2 - R_0^2}$ gravity.} \label{potential}
\end{figure}

To show the equivalence of these two frames one can use the
auxiliary fields A and B and rewrite the action (\ref{generalized})
as:
\begin{equation}
S=\frac{1}{2\kappa}\int{d^{4}x\sqrt{-g}(B(R-A)+f(A))}
\end{equation}
Varying the action with respect to $B$ we get $R=A$ which results in
generalized gravity in Jordan frame. Also varying the action with
respect to $A$ we get $B=f'(A)$ and we can write the action as:
\begin{equation}
S=\frac{1}{2\kappa}\int{d^{4}x\sqrt{-g}(f'(A)(R-A)+f(A))}
\end{equation}
Now we use the conformal transformation $g_{\mu\nu}\rightarrow
\exp(\varphi)g_{\mu\nu}$ where $\varphi=-\ln{f^{\prime}(A)}$  and
obtain the Einstein frame action (scalar-tensor gravity), as
follows:
\begin{eqnarray}
&& S=\frac{1}{\kappa^2}\int d^{4}x\sqrt{-g}[R-\frac{3}{2}[{\frac{f^{\prime\prime}(A)}{f^{\prime}(A)}}]^{2}{g^{\mu\nu}\partial_{\mu}A\partial_{\nu}A} \nonumber\\
&&-\frac{A}{f(A)^{\prime}}+\frac{f(A)}{[{f^{\prime}(A)}]^2}].
\end{eqnarray}
We can also write this action as:
\begin{equation}
S=\frac{1}{\kappa^2}\int{d^{4}x\sqrt{-g}[R-\frac{3}{2}g^{\mu\nu}\partial_{\mu}\varphi\partial_{\nu}\varphi-V(\varphi)]}
\end{equation}
where the scalar potential is:
\begin{equation}
V(\varphi)=\frac{A}{f^{\prime}(A)}-\frac{f(A)}{{f^{\prime}(A)}^{2}}
\end{equation}
For the case of $f(R)=\sqrt{R^2-R_{0}^2}$, the potential of
non-minimally coupled theory is:
\begin{eqnarray}\phi = \ln \frac{R}{\sqrt{R^{2}-R_{0}^{2}}}  ,\nonumber\\
V(\phi)=R_{0}e^{-\phi}\sqrt{1-e^{-2\phi}}. \label{pot}
\end{eqnarray}
where for convenience we changed $\varphi$ to $-\varphi$ in our
calculation. For the early universe where $R\longrightarrow \infty$,
$\phi \longrightarrow 0$ and for $R= R_0$ which is the vacuum
solution of the generalized gravity, $\phi = \ln{\sqrt{2}}$. Figure
(\ref{potential}) shows the dependence of the potential to the
scalar field and the evolution of field starts from $\phi = 0$ to
its maximum value at $\phi = \ln{\sqrt{2}}$. The final stage of the
scalar field results in de Sitter phase expansion of the universe.
\section{Instability of $f(R)$ gravity}
\label{instability} 
\begin{figure}
\includegraphics[height=3in, width= 3.5in, angle=0]{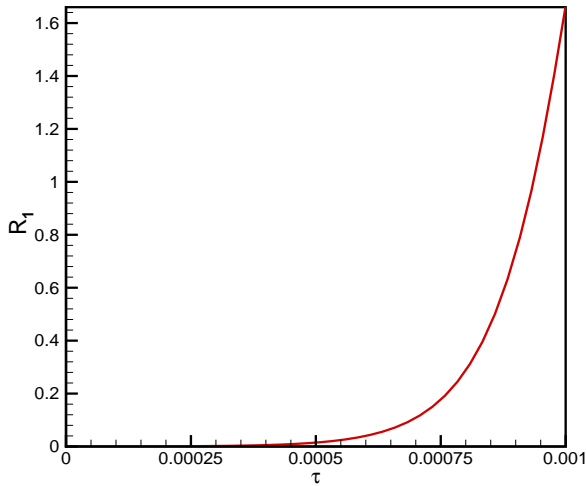}
\caption{Numerical solution of Equation (\ref{peq}) for the
perturbation of Ricci scalar around vacuum solution.}
\label{perturbation}
\end{figure}

In order to examine the instability of the
action we do a small perturbation of Ricci scalar around the vacuum
solution to see if the perturbation grows or decays. A small
perturbation around the vacuum can be written:
\begin{equation}
R=R_{v}+R_{1},
\end{equation}
where $R_{1}\ll R_{v}$ and $R_{v}$ is the Ricci scalar of the
vacuum. To find the dynamics of $R_{1}$, we expand the trace of
Equation (\ref{m_vari}) around $R_{v}$ as:
\begin{equation}
f'(R_{v}+R_{1})(R_{v}+R_{1})-2f(R_{v}+R_{1})+3\nabla^{\mu}\nabla_{\mu}f'(R_{v}+R_{1})=0,
\end{equation}
where the first order perturbation gives:
\begin{equation}
\nabla^{\mu}\nabla_{\mu}R_{1}+\frac{R_{v}f''(R_{v})-f'(R_{v})}{3f''(R_{v})}R_{1}=0.
\end{equation}
For the case of $f(R)=\sqrt{R^{2}-R^2_0}$, we have:
\begin{equation}
\nabla^{\mu}\nabla_{\mu}R_{1}+ bR_{1}=0,
\end{equation}
where $b=\frac{2\sqrt{2}}{3}R_0$. For FRW metric we can write this
equation as follows:
\begin{equation}
-\ddot{R_{1}}+a^{-2}(t)(\partial^{2}_{x}+\partial^{2}_{y}+\partial^{2}_{z})R_{1}+bR_{1}=0.
\label{perturb}
\end{equation}
Since the vacuum solution of the background results in a de Sitter
solution and exponential expansion of the universe, we use
$a(t)=\exp^{\frac{1}{2}\alpha(t-t_{0})}$ for the dynamics of the
background, in which $\alpha=(\frac{\sqrt{2}R_0}{3})^{\frac{1}{2}}$.
In the Fourier space (\ref{perturb}) can be written as:
\begin{equation}
\ddot{R}_1(k)+k^{2}e^{-\alpha(t-t_{0})}R_1(k)-bR_1(k)=0.
\end{equation}
It is clear that in the limit $t\rightarrow \infty$ we can ignore
the second term and the solution is exponentially growing. We can
also do semi-analytical calculation by using $\alpha\sim t_{0}^{-1}$
and $b\sim t_{0}^{-2}$ and as a result, the differential equation
is:
\begin{equation}
{R''}_1(k)+(e^{1-\tau}k^2-1)R_1(k)=0,
\label{peq}
\end{equation}
where prime represents derivation with respect to the dimensionless
variable $\tau=\frac{t}{t_{0}}$. For perturbations smaller than
horizon, (i.e. $k<1$) we will have exponentially growing modes,
while wavelengths larger than horizon will oscillate. Figure
(\ref{perturbation}) shows the dynamics of $R_1$ in terms of the
normalized time $\tau$ for the modes smaller than horizon.

In order to stabilize this action we can add a quadratic term to the
the action as $\alpha R^2$ and examine the stability. This stability
condition can be also obtained if one directly solves the perturbed
generalized Einstein equation:
\begin{equation}
-{R''_{1}}+e^{-(\tau-1)}(\partial^{2}_{x}+\partial^{2}_{y}+\partial^{2}_{z})R_{1}+
bR_{1}=0,
\end{equation}
where $b=\frac{2\sqrt{2}}{3}(R_0^{-1}-2\alpha)^{-1}$. For $\tau\ll1$
and in the Fourier space, one gets
\begin{equation}\ddot{R}_1(k)+(ek^{2}-b)R_1(k)=0.\end{equation}
So $R_1(k)$ would oscillate if $ek^{2}>b$, or equivalently, the
model would be stable for perturbations of all wavelengths if
\begin{equation}2R_0\alpha >1
\label{is_cond}
\end{equation}
Adding $\alpha R^{2}$ term not only makes the action stable but it
will be also important for the early time inflation. To show this we
can ignore the first term of action at the early universe while
$R>>R_0$ and consider the action as:
\begin{equation}
f(R)\simeq\alpha R^{2}.
\end{equation}
For the radiation dominant epoch where $p = 1/3\rho$, the trace of
energy momentum tensor is zero, $T^{\mu}{}_{\mu} = 0$. So the
generalized Einstein equation reduces to
\begin{equation}\nabla^\mu\nabla_\mu f'(R)=0.
\end{equation}
For homogenous universe we will have only temporal derivative of
$f'(R) = 2\alpha R$, which results in:
\begin{equation}
\nabla^{0}\nabla_{0}R=0,
\end{equation}
where for FRW metric $\Gamma^{0}_{0\lambda} = 0$ and we can replace
the covariant derivatives with partial ones which results in:
\begin{equation}R=Ct+D,\end{equation}
$C$ and $D$ can be chosen so that when $t\simeq 0$, $R\simeq D$. So
the scale factor would behave exponentially ($a(t)\propto e^{\frac
{\sqrt{D}}{2}t}$) in the very early epoch.

The same result can be achieved by looking to the instability of
scalar potential of $V(\phi)$ in conformal transformation of metric
to Einstein frame. For the case of $f(R) = \sqrt{R^2 - R_0^2}$ the
corresponding scalar potential as shown in Figure (\ref{potential})
is instable at the maximum of the potential since $V^{''}<0$. Adding
the quadratic term gives the scalar potential as:
\begin{equation} V(\phi(R))= \frac{R}{\frac{R}{\sqrt{R^{2}-R_0^{2}}}+2\alpha
 R}-\frac{\sqrt{R^{2}-R_0^{2}}+\alpha
 R^{2}}{(\frac{R}{\sqrt{R^{2}-R_0^{2}}}+2\alpha
 R)^{2}}.\end{equation}
Using the trace of generalized Einstein gravity, vacuum happens
again at $R = \sqrt{2} R_0$. The second derivation of potential at
$R=\sqrt{2}R_0$ yields:
\begin{equation}
V''( R = \sqrt{2} R_0) = \frac{-1+4R_0^2
\alpha^2}{R_0(1+2R_0\alpha)^4}.
\end{equation}
The stability condition $V''>0$, implies $2R_0\alpha>1$, which is in
agreement with (\ref{is_cond}), the approach of perturbation of
Ricci scalar.

\section{Conclusion}
\label{conclusion} Here in this work we introduced a new action for
the gravity as $f(R) = \sqrt{R^2 - R_0^2}$ which can imply a late
time acceleration for the universe. Expanding this action results in
$F(R) = R -1/2R_0^2/R + ...$ which is similar to that of $1/R$
gravity models. In the strong gravity regime, action can be written
as $f(R) = R$ and the Einstein-Hilbert action is recovered. By the
two different metric and Palatini approaches we obtained the field
equations of gravity and showed that by tuning the parameter of
model $R_0$ , universe can enter an acceleration phase in any
desired redshift.

We also obtained the equivalent theories as the Brans-Dicke and
non-minimally coupled scalar-tensor gravity. Finally the instability
of this action was examined by perturbing the Ricci scalar around
the vacuum solution which showed that we have instability for the
perturbations smaller than the horizon wavelength. Adding the
quadratic term of $\alpha R^2$ to the action stabilize it for this
type of perturbations for $2\alpha R_0>1$. We also tested the
stability condition for the scalar field in the equivalent
scalar-tensor gravity which resulted in $2\alpha R_0>1$.

\section*{Appendix}
In Section \ref{dynamics} the solution of the generalized Einstein
equation in the spherically symmetric space for the special case of
vacuum solution is obtained. We try to find another solution
perturbing Ricci scalar around the vacuum solution as:
\begin{equation}
R=\sqrt{2}R_0+ f(r)\epsilon,
\end{equation}
where $\varepsilon$ is a small dimensionless parameter. Substituting
this term in the vacuum field equation of (\ref{trace1}) and
ignoring the terms of second order in $\varepsilon$ gives:
\begin{equation}
f(r)^{''}-\frac{2\sqrt{2}}{3}R_{0}f(r)=0,
\end{equation}
where the solution is as:
\begin{equation}
f(r)=A\sin{(\sqrt{\frac{2\sqrt{2}R_0}{3}}r)}+B\cos{(\sqrt{\frac{2\sqrt{2}R_0}{3}}r)}.
\end{equation}
We can ignore the "sin" term by the argument that the maximum of
Ricci scalar is around the gravitational source which should be
vanished at infinity. So the overall Ricci scalar can be written as:
\begin{equation}
R=\sqrt{2}R_{0}+R_{0}\cos{(\sqrt{\frac{2\sqrt{2}R_0}{3}}r)}\varepsilon
\end{equation}
Substituting the Ricci scalar in (\ref{Ricci}), we obtain the metric
as:
\begin{eqnarray}
&&\lambda(r)=-\ln[1+\frac{c_1}{r}-\frac{c_2}{r^2}-\frac{\sqrt{2}}{12}R_0r^2\\
\nonumber
&&-R_{0}\varepsilon[\frac{1}{r^2}\int{r^3\cos{(\frac{1}{3}2^{\frac{3}{4}}\sqrt{3R_{0}}r)}}dr\\\nonumber
&&-r\int{r^2\cos{(\frac{1}{3}2^{\frac{3}{4}}\sqrt{3R_{0}}r)}}dr]]
\end{eqnarray}
It is obvious that if we set $\varepsilon=0$ (ignore the
perturbation term) we will achieve the solution of constant Ricci
scalar and we can determine $c_1=-2GM/c^2$ and $c_2=0$ .

{\bf Acknowledgements} We would like to thank M. M. Sheikh Jabbari
for his useful comments in this work.


\end{document}